\documentclass{ws-procs10x7}
\usepackage{balance}

\newcolumntype{d}[1]{D{.}{.}{#1}}

\def\be{\begin{eqnarray}}
\def\en{\end{eqnarray}}

\makeindex
\begin{document}

\title{Probing New Physics in $B_s$ and $D$ mixings,
 and $A_{\rm CP}(B^+ \to J/\psi K^+)$}

\author{George W.S. Hou$^*$}

\address{Department of Physics, National Taiwan
University, Taipei, Taiwan 10617, R.O.C.\\
$^*$E-mail: wshou@phys.ntu.edu.tw}

%%%%%%%%%%%%%%%%%%%%%%%%%%%%%%%%%%%%%%%%%%%%%%%%%%%%%%%%%%%%%%%%%%%%%%%%%
% You may repeat \author \address as often as necessary                 %
%%%%%%%%%%%%%%%%%%%%%%%%%%%%%%%%%%%%%%%%%%%%%%%%%%%%%%%%%%%%%%%%%%%%%%%%%

\twocolumn[\maketitle\abstract{A 4th generation could be
consistent with the recently measured $\Delta m_{B_s}$ as well as
${\cal B}(b\to s\ell^+\ell^-)$, which are SM-like, but generate
large $\sin2\Phi_{B_s} \simeq -0.5$ to $-0.7$. The sign is
determined by the hint for New Physics in CPV measurements in
charmless $B$ decays. The $4\times 4$ unitarity allows one to
connect to all processes involving flavor. Fixing $V_{t'b}$,
$V_{t's}$ and $V_{t'd}$ by $Z\to b\bar b$, $b\to s$ and $s\to d$
processes, we predict $D$ mixing to be close to the current bound.
As a further corollary, we suggest that $A_{\rm CP}(B^+ \to J/\psi
K^+)$ could be at 1\% level or higher, where we give plausibility
of an associated strong phase. Our predictions can be tested in
the near future.}
 ]

\section{Introduction: SM Reigns?}

The New York Times reported on July 4th the measurement of $B_s$
mixing at the Tevatron, stating that ``it was right on the money
as predicted by the Standard Model", and quoting a CDF
spokeswoman, ``Our real hope was for something bizarre".

The measured\cite{dmsCDF} $\Delta m_{B_s} = 17.77\pm 0.10 \pm
0.07\ {\rm ps}^{-1}$ is indeed consistent with SM, but there is
still hope for something bizarre: Can CP violation in $B_s$ mixing
be large? Given that $\sin 2\Phi_{B_s}^{\rm SM} = -\sin2\beta_s
\sim -0.04$ is very small, {\it any definite measurement at the
Tevatron would imply New Physics} (NP).
There is reason for hope. The $\Delta m_{B_s}$ value is somewhat
lower than the CKM/UT fit projections made without using $\Delta
m_{B_s}$ in the fit. %This allows for NP.

In the 4 generation model we predict $\sin2\Phi_{B_s}$ is {\it
large and negative}, with two corollaries. One is finite $D$
mixing close to current bounds, the other is observable direct CPV
(DCPV) in $B^+ \to J/\psi K^+$ decay. Mixing dependent CPV (TCPV)
measured in $B^0\to J/\psi K^0$, namely\cite{Hazumi} ${\cal
S}_{J/\psi K} = 0.685 \pm 0.032$, is also low against CKM/UT fit
predictions, which could be due to NP phase.

Admittedly, SM4 has troubles with precision EW tests.\cite{EW} But
with the LHC approaching, we should keep an open mind. For $N_\nu$
counting, as discussed by Soddu,\cite{eV4} massive neutrinos call
for NP.
The reason we focus on the 4th generation is its ease in affecting
heavy meson mixings and other electroweak penguins
(EWP),\cite{HWS} and it naturally brings in a new CPV
phase.\cite{AHphase}
%Given that flavor and CPV {\it is} the frontier in our
%search for NP, with several hints already, the 4th generation
%should not be taken lightly.

\section{Large CPV in $B_s$ Mixing}

The 4 generation unitarity for $b\to s$ transitions is $\lambda_u
+ \lambda_c + \lambda_t + \lambda_{t'} = 0$, where $\lambda_i
\equiv V_{is}^*V_{ib}$. Since $|\lambda_u| < 10^{-3}$ by direct
measurement, one effectively has
\begin{equation}
\lambda_t \cong -\lambda_c - \lambda_{t'},
% \label{lamt}
\end{equation}
where one has a NP CPV phase\cite{AHphase} through $\lambda_{t'}
\equiv V_{t's}^*V_{t'b} \equiv r_{sb}\,e^{i\phi_{sb}}$, and
Eq.~(1) becomes a triangle with potentially large area, i.e. large
CPV effect.

The formula for $B_s$ mixing is\cite{BsDmix}
\begin{eqnarray}
M_{12} &\propto& f_{B_s}^2 B_{B_s}
 \Bigl\{\lambda_c^2 \, S_0(t,t) \nonumber \\
 && - 2\lambda_c\lambda_{t'}\,\Delta S_0^{(1)}
%                               \bigl[S_0(t,t') - S_0(t,t)\bigr] %\right.
 + \lambda_{t'}^2\,\Delta S_0^{(2)}
%  \bigl[S_0(t',t') - 2S_0(t,t') + S_0(t,t)\bigr]
      \Bigr\},
 \label{M12}
\end{eqnarray}
where $S_0(t,t)$ gives SM3 top effect, and the $t'$ effects are
GIM subtracted and vanish with $\lambda_{t'}$, analogous to
$\Delta\,C_i \equiv C^{t^\prime}_{i} - C^t_{i}$
terms\cite{AHphase} that modify the Wilson coefficients $C_i$ for
$b\to s$ decays. One also has analogous strong dependence on
$m_{t'}$ is, i.e. nondecoupling of $t$ and $t'$ from box and $Z$
penguins.\cite{HWS}

Taking $m_t=170$ GeV and the central value of
$f_{B_s}\sqrt{B_{B_s}} = 295 \pm 32$ MeV from lattice, we find
$\Delta m_{B_s}^{\rm SM} \sim 24$ ps$^{-1}$, which is on the high
side compared with Eq.~(1). Of course $f_{B_s}\sqrt{B_{B_s}}$
could be lower, but {\it it could also be higher}. One may
therefore need SM4 to bring $\Delta m_{B_s}$ down a bit.

%\begin{figure}[t]
%\centerline{\psfig{file=a0_a1.eps,width=2.2in}} \caption{Lattice
%calculations of $a_0$ and $a_1$ masses as a function of $m_\pi^2$
%[6].} \label{fig1}
%\end{figure}

% For four generation
\begin{figure}[t!]
\smallskip  %\smallskip
%\vspace{-3mm}
\hspace{2mm}
 %\hspace{1mm}%\vspace{4.mm}
\begin{center}
\includegraphics[width=1.6in,height=1.1in,angle=0]{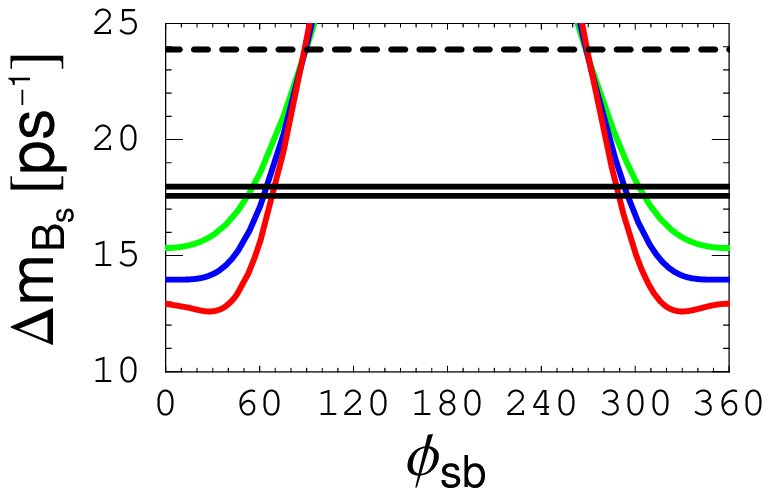}\\
\hspace{-2.5mm}%\vspace{-5.mm}
\includegraphics[width=1.6in,height=1.1in,angle=0]{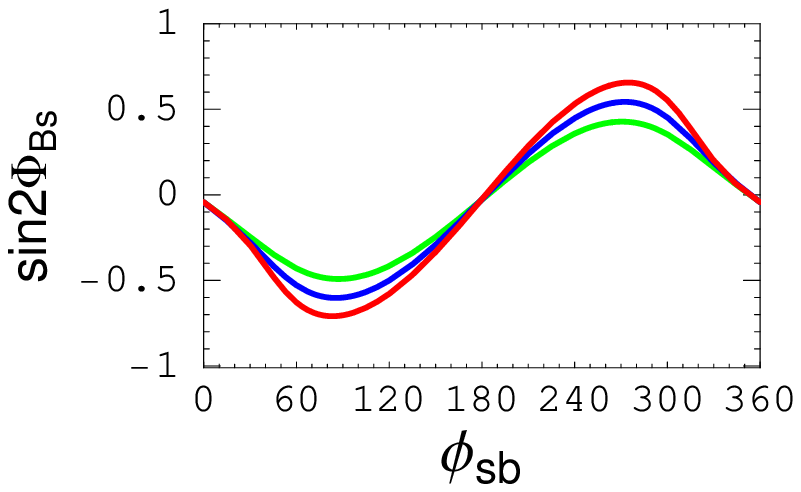}
%\vspace{6.mm}
%\smallskip\smallskip%\smallskip\smallskip\smallskip
\vspace{-2mm}
\end{center}
\caption{
 (a) $\Delta m_{B_s}$,
 (b) $\sin2\Phi_{B_s}$
 vs $\phi_{sb}$, for $m_{t^\prime}=300$ GeV and
 $r_{sb}=$ 0.02, 0.025 and 0.03.
 Larger $r_{sb}$ gives stronger variation.
}
% \label{fig:btos}
\vspace{-3mm}
\end{figure}

Keeping $f_{B_s}\sqrt{B_{B_s}} = 295$ MeV,
in Fig.~1(a) we plot $\Delta m_{B_s}$ vs $\phi_{sb}$ for $m_{t'}
=$ 300 GeV and $r_{sb} =$ 0.02, 0.025 and 0.03, where dashed line
is the SM3 value, and solid band is the 2$\,\sigma$ range of
Eq.~(1). We see that\cite{BsDmix} $\Delta m_{B_s}$ comes down to
the CDF range in 1st and 4th quadrant.
%For given $r_{sb}$, one projects a rather narrow $\phi_{sb}$ range.
For $r_{sb} =$ 0.02, 0.025, 0.03, we find $\phi_{sb} \simeq
52^\circ$--$55^\circ$, $62^\circ$--$64^\circ$,
$67^\circ$--$69^\circ$. This implies large CPV, i.e. large
$\sin2\Phi_{B_s}$, which is plotted in Fig.~1(b).
%, where $-0.04$ of SM is recovered for $\phi_{sb} = 0$.

It is important to note that the parameter range above not only
gives SM-like $\Delta m_{B_s}$, it also gives SM-like ${\cal
B}(b\to s\ell^+\ell^-)$,\cite{Kpi0HNS} as the latter is also
dominated by EWP and box diagrams.\cite{HWS}
In fact, combining $\Delta m_{B_s}$ with the $b\to s\ell^+\ell^-$
rate, $|\phi_{sb}| \gtrsim 55^\circ$ is implied,\cite{BsDmix}
which practically rules out the allowed range from $\Delta
m_{B_s}$ for $r_{sb} \sim$ 0.02. This leads to $|\sin2\Phi_{B_s}|
\sim 0.5$ to $0.7$.
%, with a second possibility of slightly weaker positive value.
Thus, things may still turn ``bizarre". Given that
CDF has made precision measurement\cite{dmsCDF} of $\Delta
m_{B_s}$, {\it can one pull off another coup in measuring
$\sin2\Phi_{B_s}$, before LHC start?} Any definite measurement
would be a discovery of NP!

Currently we have two hints for NP in CPV $b\to s$ transitions.
Interestingly, they favor $\sin2\Phi_{B_s} < 0$. One hint is TCPV
in $b\to s\bar qq$: the $\Delta {\cal S} \equiv {\cal S}_{s\bar
qq} - {\cal S}_{\bar ccs} < 0$ problem.\cite{Hazumi} The other
hint is difference in DCPV between $B\to K^+\pi^-$ vs $K^+\pi^0$:
the $-\Delta {\cal A}_{K\pi} \equiv {\cal A}_{K\pi} - {\cal
A}_{K\pi^0} < 0$ problem.\cite{Barlow}

All measurements of TCPV in $b\to s\bar qq$ modes at present give
values lower\cite{Hazumi} than charmonium modes, giving a combined
significance of 2.5$\sigma$. What aggravates this is the SM
expectation of $\Delta {\cal S} > 0$. In QCDF, it was
shown\cite{HNRS} that ${\cal S}_{K\pi}$ and ${\cal S}_{\phi K}$
are more robust than rates, which have large hadronic
uncertainties. However, ${\cal S}_{\eta'K}$ gets diluted away by
effect of the large rate. In a model independent way,\cite{SMH} it
has recently been shown that, if this discrepancy persists as data
improves, it would definitely imply NP.

The difference $\Delta {\cal A}_{K\pi} \simeq 0.15$ is now
established.\cite{Barlow} It is a puzzle because naively one
expects it to be smaller. There are two possibilities. One is an
enhancement of the color-suppressed tree ($C$). The other is from
$P_{\rm EW}$ (the EWP), which would demand NP CPV effect. The
latter case was demonstrated with the 4th
generation,\cite{Kpi0HNS} where the $\phi_{sb}$ phase of Eq.~(2)
affects $P_{\rm EW}$. The $C$ and $P_{\rm EW}$ efforts were
recently joined\cite{HLMN} and carried to NLO in PQCD
factorization. Both trends for $\Delta {\cal A}_{K\pi}$ and
$\Delta {\cal S}$ can be accounted
Interestingly, predictions for ${\cal A}_{K^0\pi^0}$ and $R_c$,
$R_n$ ratios are in good agreement with the new experimental
results,\cite{Barlow} while further prediction for ${\cal S}_{\rho
K}$ can be tested in the future.

As these are CPV measurables, the upshot from the $\Delta {\cal
S}$ and $\Delta {\cal A}_{K\pi}$ discussion is that they {\it
select} $\sin2\Phi_{B_s} < 0$ in SM4.

\section{$D$ Mixing Prediction}

Four generation unitarity links all flavor changing and CPV
processes together. With $V_{t's}^*V_{t'b}$ large, one has to
check for consistency\cite{globalHNS} with other processes. A
typical $4\times 4$ CKM matrix is given in Ref. 13. One first
saturates the $Z\to b\bar b$ bound with $V_{t'b} \simeq -0.22$,
which then fixes $V_{t's}$ by $b\to s$ effects. Applying the
stringent kaon physics constraints fixes $V_{t'd}$.

A very important test is $b\to d$ transitions. Remarkably, when
the above procedure is done, it was found that $B_d$ mixing and
associated CPV (``$\sin2\phi_1$"), as well as other $b\to d$
effects, all do not get much affected. The reason is because one
cannot easily tell apart (at present level of errors) the $b\to d$
unitarity quadrangle in SM4, from the triangle in SM3.

One striking feature of the ``fitted" $4\times 4$ matrix is that
$V_{t'd} \simeq -0.0044\,e^{-i10^\circ}$, $V_{t's} \simeq
-0.114\,e^{-i70^\circ}$, and $V_{ub'} \simeq
0.068\,e^{i61^\circ}$, $V_{cb'} \simeq 0.116\,e^{i66^\circ}$ are
not smaller than 3rd generation elements. Though somewhat
uncomfortable, this is data driven, and draws our interest to $D$
mixing, since
\begin{equation}
V_{ub'}V_{cb'}^* \equiv r_{uc}\, e^{-i\,\phi_{uc}} = +0.0033\,
e^{-i\,5^\circ},
% \label{lambp}
\end{equation}
would affect $c\to u$ transitions via $b'$ loops. Since
$|V_{ub}V_{cb}| \lesssim 10^{-4}$ by direct measurement, the
unitarity condition is effectively
\begin{equation}
V_{ud}V_{cd}^* + V_{us}V_{cs}^* + V_{ub'}V_{cb'}^* \cong 0,
% \label{unitarity_cu}
\end{equation}
with $V_{ud}V_{cd}^* \simeq -0.218$ and $V_{us}V_{cs}^* \simeq
0.215$ real to better than 3 decimal places, much like in SM.
These govern $c\to u\bar dd$ and $u\bar ss$ processes, where
especially the latter could generate width difference $y_D =
\Delta \Gamma_D/2\Gamma_D$ through long-distance effects.

Though small, $V_{ub'}V_{cb'}^*$ of Eq. (3) can affect $D^0$-$\bar
D^0$ mixing, because $m_{b'} \sim m_{t'}$ is expected, hence very
heavy. The short distance effect corresponds to the $\Delta
S_0^{(2)}$ term in Eq.~(2), with $f_{B_s}^2B_{B_s} \lambda_{t'}
\longrightarrow f_{D}^2B_{D}|V_{ub'}V_{cb'}^*|$, and $m_{t'}
\longrightarrow m_{b'}$. This generates $x_D^{\rm SD}$ which would
be vanishingly small in SM3 because of $|V_{ub}V_{cb}|^2$
suppression.

% For Dmixing
%
\begin{figure}[t!]
\smallskip  %\smallskip
%\vspace{-3mm}\hspace{-1.5mm}
\begin{center}
\includegraphics[width=1.6in,height=1.1in,angle=0]{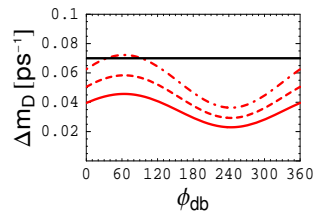}
\end{center}
\vspace{-1mm} \caption{
 $\Delta m_D$ vs $\phi_{db}$ ($\sim 10^\circ$ expected)
     for $m_{b'} =$ 230, 270 and 310 (highest) GeV
     and $r_{db} = 10^{-3}$.
     The horizontal line is the experimental bound.
 }
 \label{fig:Dmix}
\vspace{-2mm}
\end{figure}

We used\cite{globalHNS} $V_{t'd}^*V_{t'b} \equiv r_{db} \,
e^{\phi_{db}}$ to fit kaon data, and found $\phi_{db} \sim
10^\circ$ and $r_{db} \sim 10^{-3}$. For illustration, we take
$f_D\sqrt{B_{D}} = 200$ MeV and plot $\Delta m_D$ vs $\phi_{db}$
in Fig.~2, for $m_{b'} =$ 230, 270 and 310 GeV. Our scenario
predicts $x_D^{\rm SD} \equiv \Delta m_D^{\rm SD}/\Gamma_D \sim
1\%-3\%$, which lies just below the current bound\cite{Pakhlov}
(horizontal line), and could be accessible soon. We find CPV in
$D^0$ mixing to be no more than $-0.2$ level, which is consistent
with null search for CPV.

There is in fact a hint for width difference. Averaging over $D^0$
decays to CP eigenstates $K^+K^-$ and $\pi^+\pi^-$
gives\cite{Pakhlov} $y_{\rm CP} = 0.90 \pm 0.42\, \%$.
Another effort is to measure $x' = x_D\cos\delta + y_D\sin\delta$
and $y' = y_D\cos\delta - x_D\sin\delta$ in wrong-sign $D^0\to
K^+\pi^-$ decays, which could arise through mixing, or from doubly
Cabibbo suppressed decays. The current best limit comes from
Belle,\cite{Pakhlov} $|x'| < 2.7\%$ and $-1\% < y' < 0.7\%$. For
small $\delta$ this implies $y' \sim y_D \sim 1\%$ and $x$ would
be not much larger. However, for strong phase $\delta \sim
20^\circ$--$50^\circ$, $x_D$ could be several times larger than
$y_D\sim 1\%$. With an active program at the B factories and
CLEO-c, and the expectation that BESIII and LHCb would start
running in 2008, it looks promising that $x_D \sim 0.01$ to $0.03$
can be discovered soon.

\section{DCPV in $B^+ \to J/\psi K^+$}

One intriguing ``prediction" we can make is ${\cal A}_{J/\psi
K^+}\neq 0$.\cite{ApsiK}

The $B^+ \to J/\psi K^+$ is dominated by the color-suppressed
$b\to c\bar cs$ tree, while inclusion of the penguin in SM3 does
not alter the weak phase, which is $\simeq 0$. But the full
amplitude is likely carrying a strong phase $\delta$, since all
color-suppressed modes observed so far seem enhanced, with
effective underlying strong phase. Examples are $B^0 \to
D^0\pi^0$, $\pi^0\pi^0$. Although the strong phase in the latter
is still not settled, the former has a strong phase $\sim
30^\circ$ that is measured. The most relevant is $B \to J/\psi
K^*$, where angular analysis gives strong phase difference between
helicity amplitudes at order $30^\circ$.

The $t'$ effect in the $Z$ penguin brings the weak $\phi_{sb}$
phase to $P_{\rm EW}$ amplitude. Unlike the above ``hadronic"
effects that enhance $C$, the virtual $Z$ produces a small
color-singlet $c\bar c$ pair that exits without much interaction,
thereby not accumulating much strong phase. While a little hand
waving, we see that both weak and strong phases are present, the
prerequisites for DCPV.

% For four generation
\begin{figure}[t!]
\smallskip  %\smallskip
\begin{center}
%\vspace{-3mm}\hspace{-1.5mm}
%%%%%\hspace{-1mm}\vspace{4.mm}
\includegraphics[width=1.6in,height=1.1in,angle=0]{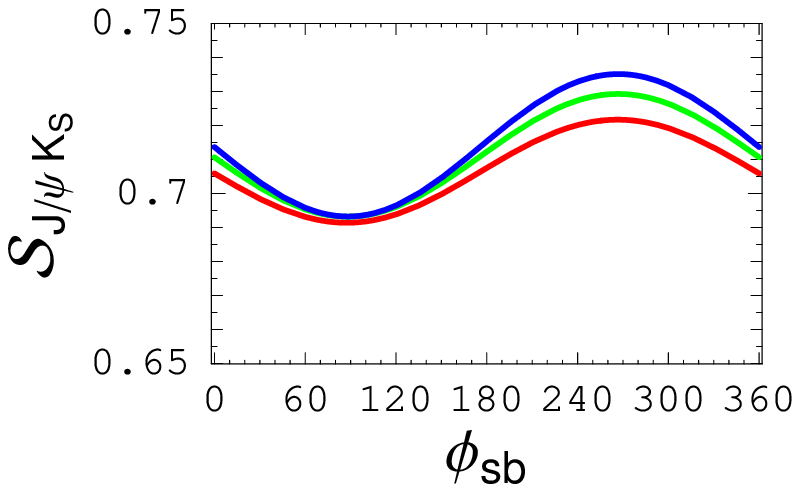}\\
%\includegraphics[width=1.6in,height=1.1in,angle=0]{tmp2}\\
%\vspace{3mm}
 \hspace{-2.5mm}
\includegraphics[width=1.64in,height=1.11in,angle=0]{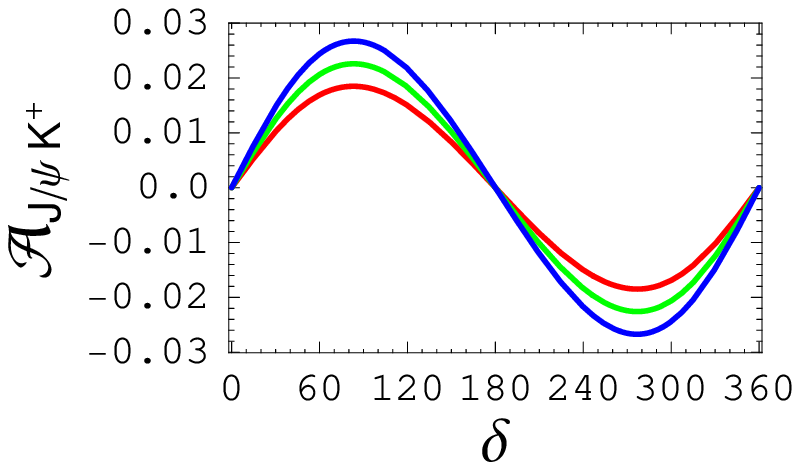}
%\vspace{-1mm}
%\smallskip\smallskip%\smallskip\smallskip\smallskip
\end{center}
\vspace{-2mm}
 \caption{(a) ${\cal S}_{J/\psi K^0}$ vs
$\phi_{sb}$, (b) ${\cal A}_{J/\psi K^+}$ vs $\delta$, for
$m_{t^\prime}=300$ GeV and $r_{sb}=$ 0.02, 0.025 and 0.03. Larger
$r_{sb}$ gives stronger variation. }
% \label{fig:fourgeneration}
\vspace{-2mm}
\end{figure}

We plot ${\cal S}_{J/\psi K}$ vs $\phi_{sb}$ in Fig. 3(a), for
$\delta = 0$. Similar to $\Delta{\cal S}$, which has $S_{J/\psi
K}$ as reference point, ${\cal S}_{J/\psi K}$ itself does dip
downwards for $\phi_{sb} \sim 65^\circ$, reaching roughly 0.69.
This does not change significantly when $\delta$ remains small.
In Fig.~3(b) we plot ${\cal A}_{J/\psi K^+}$ vs $\delta$ for
$\phi_{sb} = 65^\circ$. We find that\cite{ApsiK} {\it ${\cal
A}_{J/\psi K^+}$ can reach above 1\% for $|\delta| \sim
30^\circ$}.

The experimental situation\cite{PDG} is interesting. From ${\cal
A}_{J/\psi K^+}\sim +0.03$ based on 89M $B\bar B$s, BaBar flipped
sign by adding 35M, becoming $-0.030 \pm 0.014 \pm 0.010$, with
larger systematic error, and is now consistent with Belle value of
$-0.026 \pm 0.022 \pm 0.017$ based on 32M. The current world
average is $-0.024 \pm 0.014$, based on 166M $B\bar B$s. But the
world has now over 1000M $B\bar B$s and growing, thus, our 1\%
projection can be seriously probed. Note that the number could be
higher,\cite{ApsiK} e.g. in the less constrained $Z'$ model. To
realize a 1\% measurement, it seems that one needs to work hard on
systematic error. But this should be worthwhile if one wants to
enter the ``Super B factory" era, with 100 times more data, where
any measurement of interest is likely to be systematics limited.

With luck, our prediction can be confirmed by 2008.

%\section{Conclusion}

\section*{Acknowledgments}
I thank Makiko Nagashima and Andrea Soddu for collaboration.

\balance

\appendix

\end{document}